\documentclass[aps,prl,preprint,english,superscriptaddress,preprintnumbers,amsmath,amssymb,showpacs]{revtex4-1}
\usepackage[T1]{fontenc}
\usepackage[latin9]{inputenc}
\usepackage{amsmath}
\usepackage{graphics}
\usepackage{graphicx}
\usepackage{epsfig}
\usepackage{amssymb}
\usepackage{dcolumn}% Align table columns on decimal point
\usepackage{bm}% bold math
\usepackage{float}
\usepackage{leftidx}
\makeatletter
\@ifundefined{textcolor}{}
{%
 \definecolor{BLACK}{gray}{0}
 \definecolor{WHITE}{gray}{1}
 \definecolor{RED}{rgb}{1,0,0}
 \definecolor{GREEN}{rgb}{0,1,0}
 \definecolor{BLUE}{rgb}{0,0,1}
 \definecolor{CYAN}{cmyk}{1,0,0,0}
 \definecolor{MAGENTA}{cmyk}{0,1,0,0}
 \definecolor{YELLOW}{cmyk}{0,0,1,0}
 }

\makeatother

\usepackage{babel}

\begin{document}
%\preprint{APS/123-QED}

%\title{Stabilizing proton beam acceleration by exposing a two-specie ultra-thin foil to a circularly polarized laser pulse}

\title{Stable laser-driven proton beam acceleration from a two-specie ultra-thin foil}

\author{T.~P. Yu}
%\thanks{Electronic mail: Tongpu.Yu@tp1.uni-duesseldorf.de}
 \affiliation{Institut f\"ur Theoretische Physik I, Heinrich-Heine-Universit\"at D\"usseldorf,
40225 D\"usseldorf, Germany}
\affiliation{Department of Physics, National University of Defense Technology,
Changsha 410073, China}

\author{A. Pukhov}
 \thanks{Electronic mail: pukhov@tp1.uni-duesseldorf.de}
 \affiliation{Institut f\"ur Theoretische Physik I, Heinrich-Heine-Universit\"at D\"usseldorf,
40225 D\"usseldorf, Germany}

\author{G. Shvets}
 \affiliation{Univ Texas Austin, Dept Phys, Austin, TX 78712 USA}
 \affiliation{Univ Texas Austin, Inst Fus Studies, Austin, TX 78712 USA}

 \author{M. Chen}
\affiliation{Accelerator $\&$ Fusion Research Division, Lawrence Berkeley National
Laboratory, Berkeley, California 94720, USA}

\date{\today}

% It is always \today, today,
%  but any date may be explicitly specified

\begin{abstract}
By using multi-dimensional particle-in-cell simulation, we present a new regime of stable proton beam acceleration which takes
place when a two-specie shaped foil is illuminated by a circularly polarized laser pulse. It is observed
that the lighter protons are nearly-instantaneously separated from
the heavier carbon ions due to the charge-to-mass ratio
difference. The heavy-ions layer extensively expands in space and acts to buffer the proton layer
from the Rayleigh-Taylor-like (RT) instability that would have
otherwise degraded the proton beam acceleration. A simple three-interface model is formulated to qualitatively explain the stabilization of
the light-ions acceleration. Due to the absence of the
RT-like instability, the produced high quality mono-energetic proton bunch can be well maintained even after the laser-foil interaction concludes.
\end{abstract}


% PACS, the Physics and Astronomy Classification Scheme. %%

\maketitle
In the past decades, plasma-based ion accelerators have attracted a
lot of attention due to their potential applications for particle
acceleration, medical therapy~\cite{s1}, proton imaging, and inertial
confinement fusion~(ICF)~\cite{s2}. One of the most important issues
is the development of laser-driven protons for radiation therapy of
deep-seated tumours~\cite{s3}. Numerous experimental and theoretical
studies have been devoted to producing such proton
beams~\cite{s4,s5,s7}. However, their qualities such as beam
collimation, energy spread ($\sim 20\%$), and peak energy ($\sim 58$
MeV), are still unsatisfactory~\cite{s5}.

Recently, with the rapid development of the laser technology,
ultraintense ultrashort ultraclean (3U) laser pulse and ultrathin
 solid target have been extensively exploited to investigate the ion
 acceleration. One of the most straightforward acceleration
 mechanisms, radiation pressure acceleration (RPA)~\cite{s7}, is being
 revisited. The first RPA experiment~\cite{s8} has shown that both the
 beam quality and the energy conversion efficiency are greatly
 improved. However, the ultrathin foil is very susceptible to the
 transverse instabilities~\cite{s9}, similar to Rayleigh-Taylor-like
 (RT) instability in ICF. It sets in at the very beginning of
 laser-foil interaction and develops at the unstable interface at the
 rate of a few laser cycles~\cite{s10}. Gradually, the surface of the
 foil becomes corrugated by the laser radiation and the entire target
 is torn into many clumps and bubbles~\cite{s9}. The final energy
 spectrum of the ions shows a quasi-exponential decay with sharp
 cut-off energy. Unlike the electron acceleration in the bubble
 regime~\cite{s11}, a stable proton beam acceleration in the realistic
 three-dimensional (3D) geometry has not been demonstrated either
 theoretically or experimentally.

In this Letter, we report on a new regime of stable proton
acceleration, where a two-specie ultra-thin shaped foil is
illuminated by a circularly polarized (CP) laser pulse. We assume
the heavier (lighter) ions to be carbons (protons), respectively.
Particle-in-cell (PIC) simulations indicate that the RT
instability only causes the spreading of the carbon ions. The
protons, which are rapidly separated from the carbon ions, are
"buffered" by the carbon ion cloud by riding on the stable
proton-carbon interface. We demonstrate that, even though the
RT-unstable carbon-vacuum interface is strongly deformed, the
feed-through of the RT instability into the RT-stable
proton-carbon interface is small. Due to the absence of the RT
instability, the compact proton layer remains well-collimated even
after the laser-foil interaction concludes. In order to
elucidate the detailed acceleration process, we first describe the
results of 1D simulations. Discussion of the influence of the RT
instability on the ion acceleration follows, backed up by 3D simulations.

When a relativistic laser pulse illuminates a two-specie foil with
thickness of a few wavelengths, a collisionless shock wave is often
excited and can efficiently accelerate the ions to high
energies~\cite{s12}. With the decreasing foil thickness, the laser
radiation pressure competes with the shock wave and becomes strong
enough to push the entire foil forward. As a result, the foil
acceleration is dominated by the RPA. The critical foil thickness can
be approximately estimated by~\cite{s13}

\begin{equation}
L \sim \frac{a}{\pi} \frac{n_{c}}{n_{e}}\lambda, \label{eq:thickness}
\end{equation}

\noindent where $a=eE_{L}/m_{e}c\omega$ is the dimensionless laser amplitude, $m_{e}$ the electron mass, $n_{e}$ the electron density, and $n_{c}$ is the critical plasma density. $c$, $\omega$, $\lambda $, and $E_{L}$ are the light speed in vacuum, the laser frequency, wavelength and electric field, respectively. In the 1D RPA model, the target motion equation is governed by

\begin{equation}
\rho\frac{d(\gamma\beta)}{dt}=\frac{{E_{L}}^{2}}{2\pi
c}\frac{{1-\beta}}{{1+\beta}}, \label{eq:rpa}
\end{equation}

\noindent where $\rho=\sum_{\substack{i} }m_{i}n_{i}L$ is the target
area mass density, $m_{i}$ and $n_{i}$ are the ion mass and density,
$\beta=v/c$ is the target velocity and $\gamma=1/\sqrt{1-\beta^{2}}$ is the relativistic factor.
We can see that the target dynamics
is defined by the area density, not the detailed foil composition. In
principle, the heavier ions can be efficiently accelerated to the same
velocity as the lighter protons and electrons.

We simulate the described mechanism using the PIC code VLPL~\cite{s14}. The longitudinal length of the 1D simulation
box is $x=60\lambda$ sampled by $6\times10^{4}$ cells, enough to
resolve the expected density spikes. Each cell contains $100$
numerical macro particles in the plasma region. The target is $0.1\lambda$
thick, located at $x=10\lambda$ and composed of carbon ions and
protons with the same number density $46.7n_{c}$, which gives the electron
density $n_{e}=320n_{c}$. A CP laser pulse with the wavelength
$\lambda=1.06\mu m$ is incident on the target from the left
boundary. The wave front of the laser arrives at the target surface at
$t=10T_{0}$, where $T_{0}=\lambda/c$ is the laser cycle. The laser
pulse is homogeneous in space but has a trapezoidal profile (linear
growth - plateau - linear decrease) in time. The duration is
$\tau_{L}=10T_{0}$ ($1T_{0}-8T_{0}-1T_{0}$). The dimensionless laser amplitude $a=100$ is chosen to satisfy Eq.~\eqref{eq:thickness}.

\begin{figure}[!htb]
\suppressfloats\includegraphics[width=9cm]{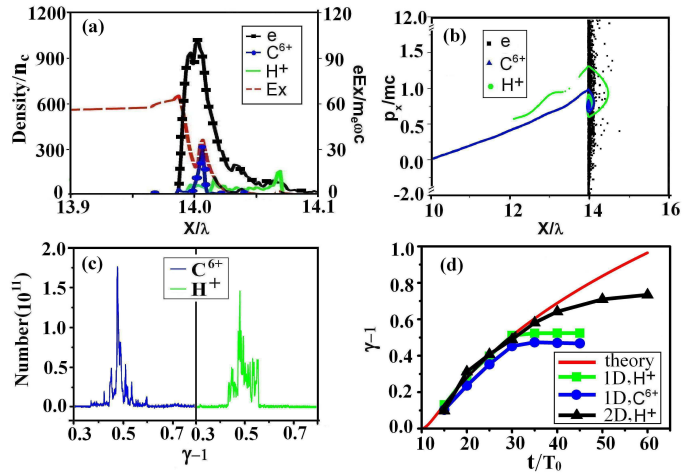}\caption{\label{f1}(Color
online). (a) Density distribution of the electrons (black), protons
(green), and carbon ions (blue) at $t=20T_{0}$. The red curve shows
the electric field $E_{x}$, which is normalized to
$E_{0}=m_{e}c\omega/e=3.2\times10^{12}V/m$. (b) The corresponding
phase space distribution at $t=20T_{0}$.  (c) Energy spectrum of the
carbon ions (blue, dark) and protons (green, light) at
$t=30T_{0}$. (d) Energy evolution in time from PIC simulations and the
1D theory. The laser pulse is incident from the left side and touches
the foil at $t=10T_0$.}
\end{figure}

Figure \ref{f1}(a) shows the particle density distribution at
 $t=20T_{0}$. In the initial stage, the
 laser pressure is transferred to the electrons, resulting in the
 charge separation~\cite{s13}. Because carbons and protons are initially co-located, the
 protons experience a higher acceleration due to their higher charge
 to mass ratio ($Z_{i}/m_{i}$). The time for protons to separate from
 the carbon ions is approximately $t_{sep}=\sqrt{2Lm_{H}/eE_L}=
 2.5fs$, which is so short that can be considered instantaneous.
 Later on, the two ion species start experiencing very different acceleration field, as shown
 by the red curve in Fig.~\ref{f1}(a). The considerably higher electric field inside the carbon layer compensates for carbon's lower $Z_{i}/m_{i}$ ratio enabling them to catch up with the protons. Eventually, both species travel together, without separating any further. The entire foil acceleration proceeds until the end of the laser-foil interaction at
 $t=35T_{0}$. Fig.~\ref{f1}(b) exhibits the phase space
 distribution. We can see that the carbon ions fall back behind the protons, accompanied by a long low-density tail. The fact that both
 ions show an obvious "spiral structure"~\cite{s15} in phase space
 provides a direct evidence for the acceleration process described
 above.

The ion energy spectrum is shown in Fig.~\ref{f1}(c). At $t=30T_{0}$,
the peak energy of the carbon ions is up to 480MeV/u. For
protons, all of them are accelerated to high energies although the
energy spectrum is somewhat wider. Fig.~\ref{f1}(d) plots the ion
energy evolution. Here, we make use of the averaged energy for both
species. At $t=35T_{0}$, the laser-foil interaction is over so that
the ion energy doesn't increase any more. Overall, the observed ion
acceleration in the 1D simulations is consistent with the predictions of
the 1D theory of Ref.~\cite{s13}.

However, multi-dimensional simulations exhibit a radically different
 acceleration dynamics because multi-dimensional effects - such as
 transverse expansion of the bunch and the RT instability - come into
 play. In order to extend the 1D model to the 2D simulations smoothly, we
 employ a shaped foil target (SFT)~\cite{s17} to
 compensate for the transverse profile of the laser pulse. Taking the
  Gaussian laser for example, the foil thickness should be
 matched transversely by the Gaussian function
 $L=max[L_{max}exp(-y^{2}/\sigma_{T}^{2}), L_{cut}]$, where $L_{max}$
 is the maximal foil thickness, $L_{cut}$ the cutoff thickness, and
 $\sigma_{T}$ the spot radius. In the following 2D case, the
 simulation box is $X\times Y=80\lambda\times32\lambda$, sampled by
 $16000\times400$ cells. Each cell contains $100$ macro particles in the
 plasma region. The foil is initially located at $x=10\lambda$ with
 parameters $L_{max}=0.1\lambda$, $L_{cut}=0.05\lambda$, and
 $\sigma_{T}=7\lambda$. The carbon ion density is $51.9n_{c}$,
 intermingled with protons of the density $8.64n_{c}$ so that the
 total electron density is $320n_{c}$. A Gaussian laser pulse with the
 focal size $\sigma_{L}=8\lambda$ is incident from the left
 boundary. All other parameters are the same as in the 1D case.

\begin{figure*}[!htb]
\suppressfloats\includegraphics[width=15.0cm]{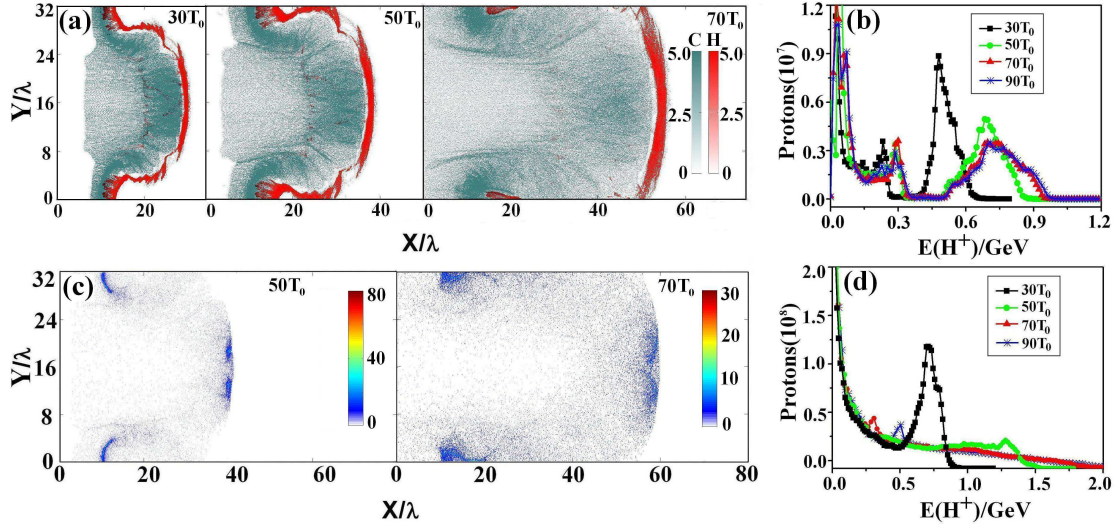}\caption{\label{f2}(Color
online). (a) Contours of protons (red, dark) and carbon ions (cyan,
light) in the 2D case at different time points: $t=30T_{0},50T_{0}$,
and $70T_{0}$. The colorbar represents the proton numbers log(N). (b)
Proton energy spectrum at: $t=30T_{0}$ (square, black), $t=50T_{0}$
(circle, green), $t=70T_{0}$ (triangle, red), and $t=90T_{0}$ (star,
blue). For comparison, Frame (c) shows the proton density distribution
in a pure Hydrogen foil and Frame (d) corresponds to its energy
spectrum evolution.}
\end{figure*}

\begin{figure}[!htb]
\suppressfloats\includegraphics[width=9cm]{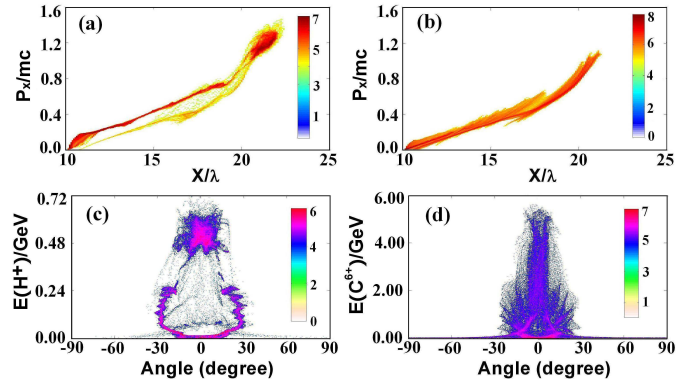}\caption{\label{f3}(Color
online).  Phase space of (a) protons and (b) carbon ions at
$t=30T_0$. An obvious spiral structure is observed only for
protons. Frame (c) and (d) show the carbon ion and proton energy
distribution as a function of the divergency angle at $t=30T_0$.}
\end{figure}

Fig.~\ref{f2}(a) shows the space distribution of the carbon ions and
 protons at different times. In each frame, the cyan color marks the
 carbon ions and the red color shows the protons. Obviously, the
 carbon ions behave totally different as compared with the 1D
 simulations. They spread widely in space and do not form a compact
 mono-energetic bunch. On the contrary, the protons from
 the center part of the foil always ride on the carbon ion front and
 form a compact bunch. The sharp front separating the species is well
 defined and remains stable even after the laser-foil interaction concludes.
  We can get further understanding of the acceleration process from the phase
  space distribution in Fig.~\ref{f3} (a) and (b). On the one hand, the carbon ions evolve into a wide cloud in
 space. On the other hand, their front co-moves with the fast protons
 so that the gap between the two species is always small. The protons
 show a clear spiral structure, like a "matchstick", which coincides
 with the 1D simulation result in Fig.~\ref{f1}(b). We don't observe
 any obvious transverse instability in the compact proton
 layer. Fig.~\ref{f2}(b) shows the proton energy spectrum. As
 expected, the peak is well pronounced and the dispersion is
 suppressed. The peak energy evolution is presented in
 Fig.~\ref{f1}(d), which is also in accordance with the 1D RPA
 model. Fig.~\ref{f3}(c, d) plots the ion energy-divergency
 distribution at $t=30T_{0}$. The high quality proton bunch with the
 energy $\sim500MeV$ and the opening angle $\sim5.5^{\circ}$ forms and
 persists in time even after the laser-foil interaction concludes.

\begin{figure}[!htb]
\suppressfloats\includegraphics[width=9cm]{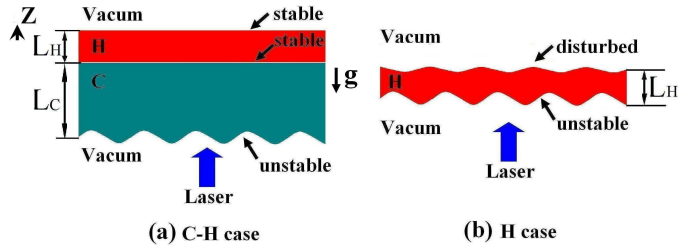}\caption{\label{f4}(Color
online).  Schematic of laser-foil interaction in (a) C-H case and (b)
pure H case. The red color marks the protons and the cyan represents
the carbons. In case (a), there are three interfaces: carbon-vacuum,
carbon-proton, and proton-vacuum. Only the first interface is
unstable. In case (b), both interfaces are finally unstable.}
\end{figure}

The stability of the proton acceleration in the 2D simulations can be
attributed to two effects. Firstly, the protons completely separate
from the carbon ions and form a thin shell, which is a prerequisite
for the stable proton acceleration. Such a separation of the ion
species can be understood within the 1D formalism developed in
Ref.~\cite{s13} and the 1D simulation above. Secondly, the heating of the carbon ions forms
 an extended cloud that prevents short-wavelength perturbations from feeding through into the
thin proton shell. We can use a simple three-interface model
as shown in Fig.~\ref{f4} to explain the stabilization.
It is helpful to consider the problem from the classical RT
instability~\cite{s18} which occurs when a light fluid is accelerated
into a heavy fluid. In the accelerating reference frame of the foil, the
perturbation pressure $p$ satisfies:

\begin{equation}
\frac{\partial^2}{\partial z^2}\delta p=-k_{\rm RT}^2\delta
p,\label{eq:pressure}
\end{equation}

\noindent where $k_{\rm RT}$ is the wavenumber of the RT-unstable mode. Noting that
$\delta p$ is discontinuous across the unperturbed boundary, we obtain
a solution $\delta p=A_i e^{-k_{\rm RT}z}+B_i e^{k_{\rm RT}z}$ away from interfaces,
with $A_i$ and $B_i$ being the amplitude coefficients of the perturbation inside the layer consisting of the $i$'th species. In
our case, both species have two interfaces: one with vacuum and one
with the other specie. For carbon ions ($i=C$), the only unstable interface is
the carbon-vacuum boundary, where the laser pulse interacts directly
with the carbon plasma. We derive from this model that the amplitude
of the RT instability is exponentially decaying away from the unstable
interface:

\begin{equation}
\frac{A_{H}}{A_{C}}\sim e^{-k_{\rm RT}L_{C}},\label{eq:amplitude}
\end{equation}

\noindent where $L_{C}$ is the thickness of the carbon ion layer. In the
simulations, $L_{C}$ is several times longer than $L_{H}$ so that the
long-wavelength perturbation in the carbon layer would take much more
time to grow (recall that the the growth rate of the RT instability $\gamma \propto \sqrt{g/\lambda_{RT}}$, where $g$ is the target's acceleration and $\lambda_{RT}$ is the perturbation wavelength). The feed-through from the unstable carbon-vacuum interface to the proton layer is exponentially  attenuated according to Eq.~(\ref{eq:amplitude}). This simple qualitative argument explains the stability of the sharp
carbon-proton interface. For the thin proton layer, it is also
stable because the protons are much lighter than the carbon ions. Eventually, the entire proton layer is free from
the RT instability. Besides, we believe that the small transverse size
of the foil also benefits the stabilization of the proton acceleration
in this case. Approximately, the minimal perturbation wavenumber can
be estimated by $k_{\rm RT}=2\pi /Y$, where $Y$ is the transverse foil
size. At $t=30T_0$, the carbon shell thickness is
$L_{C}\approx10\lambda$ and $Y=32\lambda$. Eq.~\eqref{eq:amplitude}
already indicates a considerable suppression of the perturbation
feed-through.

Now we compare the stable multi-component foil case with the pure
proton foil case, where the RT instability
is obvious. We again employ a matched SFT. All the
parameters are the same as above except now
$n_{H}=320n_{c}$ and the carbon ions are absent. Fig.~\ref{f2}(c) shows the proton density
distribution in space. We can see that the foil disrupts gradually and
two proton bunches with a lower density valley in the middle
form. This is very characteristic for the RT instability driven by the
laser radiation. According to the linear stability theory for the
accelerated foil~\cite{s9}, the growth time of the perturbation in the
relativistic limit can be derived as following

\begin{equation}
\frac{\tau_{RT}}{T_{0}}=\frac{\sqrt{2}}{6}\sqrt{\frac{m_{e}}{m_{i}}\frac{n_{c}}{n_{i}}\frac
{\lambda}{L}}(\frac{\lambda_{RT}}{\lambda})^{3/2}a,
\label{eq:instability}
\end{equation}

\noindent Taking into account
$\lambda_{RT}\simeq\sigma_{L}=8\lambda$ and $L=0.1\lambda$ in our
case, we estimate that the time scale of the instability should be
$2.2T_{0}$. Such a short-wavelength perturbation grows very fast so
that it reaches the other side of the foil soon. Finally, both
interfaces are unstable and the entire target collapses
quickly. Fig.~\ref{f2}(d) shows the proton energy spectrum. Although an
energy peak is observed initially, it lowers gradually and disappears
at $t=45T_{0}$, leaving a quasi-exponential spectrum. In fact, most
single-ion foils in the RPA regime show a similar simulation
result~\cite{s7,s9,s15,s17}. The main issue is the fast growth of the
short-wavelength perturbation at the unstable interface.

In order to check the robustness of the stable regime, we perform 3D
simulations while keeping all the parameters same as in the 2D
case except $\sigma_T=6\lambda$. A stable structure of the proton beam
acceleration is also observed, which indicates that the regime
described above can significantly stabilize the proton beam
acceleration in the realistic three dimensional geometry.

In conclusion, we present a new regime of stable proton acceleration
driven by the laser radiation. In this regime, we smoothly extend the
1D RPA model to multi-dimensional cases by using a two-specie
ultra-thin SFT. PIC simulations show that the transverse instability
degrades only the carbon ion acceleration and spreads them in
space. The sharp front separating the species is always stable so that
the proton layer is free from the effects of the RT
instability. Benefiting from the superpower lasers such as HiPER and
ELI, this stable regime might open a way to high quality proton beam
generation in the future.

%\begin{acknowledgments}\suppressfloats
% Note for American journals do not put the title of people anywhere in the manuscript.
We thank the fruitful discussions with F.~Q. Shao, N. Kumar,
and Y.~Y. Ma. This work is supported by the DFG programs GRK1203
and TR18. TPY thanks the scholarship awarded by China Scholarship
Council (CSC NO. 2008611025) and the NSAF program (Grant
No. 10976031).
%\end{acknowledgments}

%\suppressfloats
% Note: do not put a dot after the journal name unless when the last word
% is an abbreviation!

\end{document}